\documentclass[]{spie}  

\usepackage{amsmath,amsfonts,amssymb}
\usepackage{graphicx}
\usepackage[colorlinks=true, allcolors=blue]{hyperref}

\title{Low cost multi-purpose balloon-borne platform for wide-field imaging and video observation}

\author[a]{Francisco Oca\~na}
\author[b]{Alejandro S\'anchez-de-Miguel}
\author[c]{Aitor Conde}
\author[c]{Daedalus Team}
\affil[a]{Dep. Astrof\'isica y CC. de la Atm\'osfera, Universidad Complutense de Madrid, Av. Complutense s/n, E-28040, Madrid, Spain}
\affil[b]{Instituto de Astrof\'isica de Andaluc\'ia, Glorieta de la Astronomía, s/n. E-18008, Granada, Spain}
\affil[c]{Asociaci\'on  AstroINNOVA,  C/Poeta  Mohammed  Iqbal,  6  1-4, 14010 C\'ordoba, Spain, astroinnova@astroinnova.org}

\authorinfo{Further author information: (Send correspondence to F.O.)\\F.O.: E-mail: fog@astrax.fis.ucm.es\\  A.SdM.: E-mail: asanchez@iaa.es \\ Daedalus Team/Astroinnova: E-mail: astroinnova@astroinnova.org }

\begin{document} 
\maketitle

\begin{abstract}

Atmosphere layers, especially the troposphere, hinder the astronomical observation. For more than 100 years astronomers have tried observing from balloons to avoid turbulence and extinction.  New developments in card-size computers, RF equipment and satellite navigation have democratised the access to the stratosphere.

As a result of a ProAm collaboration with the Daedalus Team we have developed a low-cost multi-purpose platform with stratospheric balloons carrying up to 3 kg of scientific payload. The Daedalus Team is an amateur group that has been launching sounding probes since 2010. Since then the first two authors have provided scientific payloads for nighttime flights with the purpose of technology demonstration for astronomical observation.

We have successfully observed meteor showers (Geminids 2012, Camelopardalis 2014, Quadrantids 2016 and Lyrids 2016) and city light pollution emission with image and video sensors covering the 400-1000nm range.

\end{abstract}


\section{INTRODUCTION}
\label{sec:intro}  

Since the end of the XVIII\textsuperscript{th} century astronomers have been involved in balloon-borne observations. The first observations were casual stargazing without instruments. And therefore one of the first phenomena to be observed were meteors and fireballs. The first account of meteor sighting is as early as 1807\cite{forster1832annals,1999JIMO...27...45B}, while the first observational campaigns were organised for the forecasted storm of the Leonids 1899\cite{1999JIMO...27...45B}. Almost one hundred years later, the Leonids 1998 outburst was observed from a balloon with modern instruments\cite{jenniskens1998preparing}. Apart from ballon-borne campaigns, meteor storms have been studied in several airborne campaigns\cite{murray1999airborne,vaubaillon20152011}, but the cost is some orders of magnitude larger than low-cost ballon flights. 

Daedalus Team started as an amateur high-altitude balloons  launching team, for outreach and technology development in the frame of Do-It-Yourself projects. The authors FO and ASdM proposed to add astronomical payloads for nighttime missions for their research areas (meteors and light pollution) in the Astrophysics and Atmospheric Sciences Department of the Universidad Complutense de Madrid. The data of some of these missions were used in ASdM thesis\cite{sanchez2015variacion}.  

In the last 6 years we have carried out a total of 9 nighttime missions. We present the current status and the evolution path followed, driven by science and technology development, and based on COTS components.

\section{Description of the platform}
\label{sec:description}  

It is a passive craft, with basic stabilisation and 3-hour length nominal mission. Current design has up to 4 observational ports, three at the sides of the square-shaped probe and one aiming at nadir for Earth Observation. The structure is a styrofoam box with an anchorage point for the balloon/s. The probe uses one or two meteorological sounding balloons to get out of the troposphere. The probe usually rises to 30km of altitude and then falls free with a parachute. The probe has GSM, RF and GPS beacons and some missions had also bidirectional communication. During the flight the probe is chased from ground using the beacons in order to recover it as all the data is stored locally.

\begin{figure} [ht]
   \begin{center}
   \begin{tabular}{c} 
   \includegraphics[height=5cm]{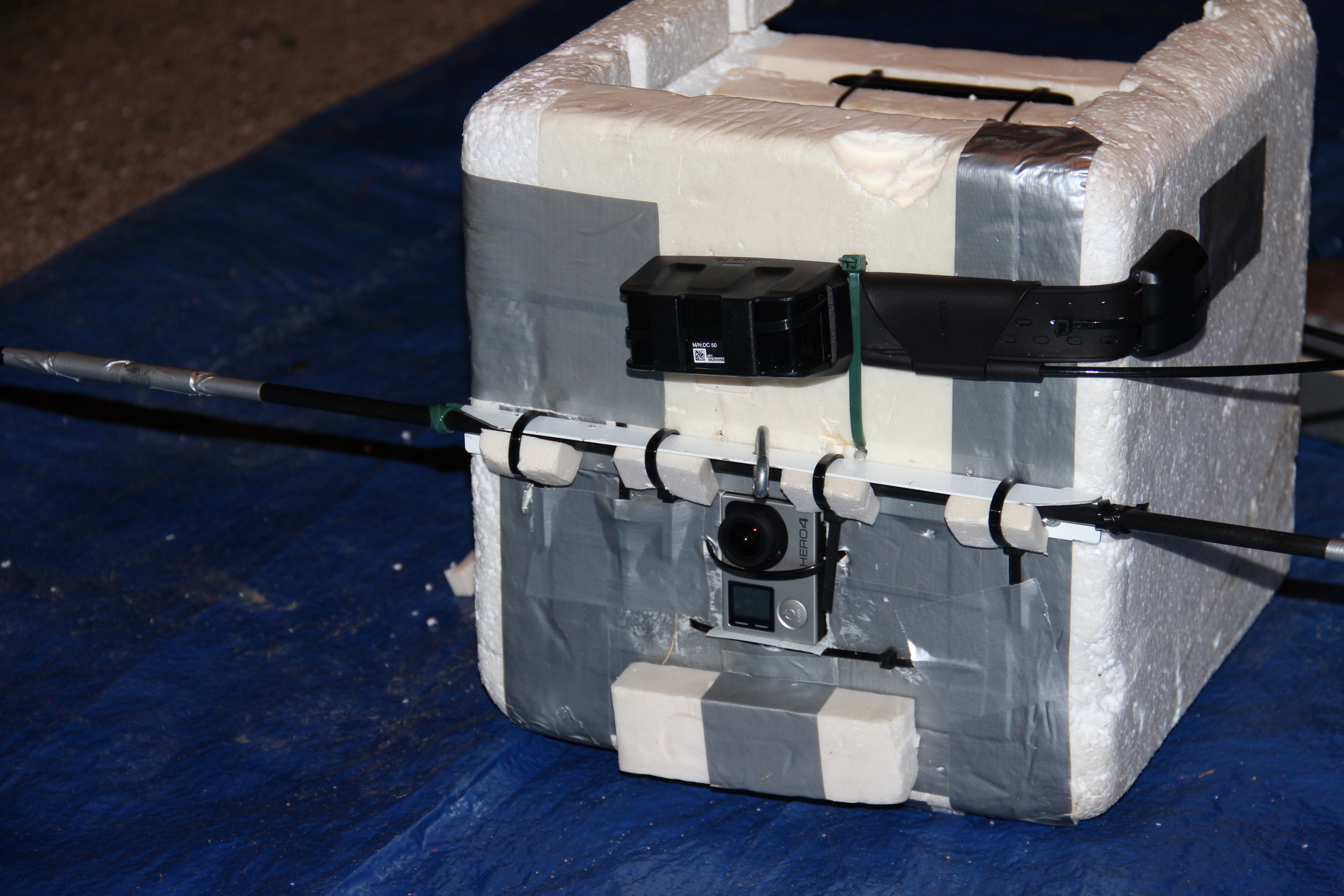}
   \end{tabular}
   \end{center}
   \caption[daedalus19] 
   { \label{fig:daedalus19} Detail of the bottom part of the probe for Daedalus 19 mission. The nadir port hosts a GoPro camera for light pollution research. Photo credit: Phillip Maier.}
   \end{figure} 

The platform is equipped with several sensors that provide information about the flight and environment. Accelerometers, magnetometers, thermometer, hygrometer characterise the flights. These values constrain the characteristics of the experiments that are qualified to flight. 

The analysis of the data from accerelometers in the 3 axis permits us to identify different flight phases: launch and troposphere with average acceleration up to 3 m/s\textsuperscript2. The flight in the stratosphere is very stable, with an average of 1 m/s\textsuperscript2 for the 3 of axis. While falling the probe is not stabilised at all, registering average values over 4-5 m/s\textsuperscript2 in all of the axis. Maximum values registered in the missions are up to 15 m/s\textsuperscript2.

The temperature is also a constraint for the instruments, for the power budget of the mission and length of the flight. The probe is not pressurised, consequently the air flows out during the ascent. Therefore the temperature decreases slowly until the burst of the balloon. During the fall the cold air flows in and batteries voltage drops significantly, but they have never frozen. 

Humidity has never become a problem. Optical instruments have shown some water drops after going through clouds, but they have evaporated soon after. Only one launch has been performed under the rain. However the lenses were not fogged after some minutes over the clouds. Nevertheless the water affects the latex and the balloon went off prematurely for that mission.

\section{Mission stages and systems operation description}
Every mission comprises several stages when part or all of the subsystems are working. In this section we describe the fundamental parts of the mission, the operation of the systems involved and the road we have taken in the development towards the final platform design.

This work comprises the results from a total of 9 nighttime missions between 2010 and 2016. All of them were recovered, and the scientific programmed was completed or partially completed in 7 of them.  

\begin{table}[ht]
\caption{Nighttime missions from 2010 to 2016. The internal code name refers to the configuration of the probe. We evaluate the scientific mission and recovery of the probe. Participants were always Daeadalus Team (D), the authors and other participant from the University (UCM), Fernando Ortu\~no (FO) and the Instituto de Astrof\'isica de Andaluc\'ia (IAA, in the frame of the Pathfinder for ORISON project). The abbreviations used are: D\#, for Daedalus mission number; Rb: radiobeacon, V: Vaisala radiosonde, DC50: Garmin DC50 tracker, S: Spot 3 GPS tracker, LP: light pollution, Rec:recovery, Partic: participants. For the equipment W:Watec 902H2, Go4: GoPro 4, Dra: Draconids, Gem: Geminids, Qua: Quadrantids, Lyr: Lyrids, Rec: recorder.} 
\label{tab:missions}
\begin{center}    
\begin{tabular}{|l|l|l|l|l|l|l|l|l|c|}
\hline
Date   & Name              & Probe                 & Mission              & Success  & Rec & Partic & Equipment                   & Stab & \multicolumn{1}{l|}{Ball} \\ \hline
2010/09/18 & D2          & NS1b                  & Noct/LP         & Yes      & Yes         & D+UCM         & Cam/Rec & No             & 2                           \\ 
2011/10/08 & D4          & Radiobeacon           & Dra          & No      & Yes         & D+UCM         & W/PC                 & No             & 2                           \\ 
2011/10/14 & D5          & NS2b                  & Noct/LP         & Partial & Yes         & D+UCM         & Mini/LX7           & No             & 1                           \\ 
2012/08/24 & D11         & NSlite                & Noct/LP         & Partial & Yes         & D             & Gopro                    & No             & 1                           \\ 
2012/12/12 & D12         & NSlite                & Gem           & Yes      & Yes         & D+UCM         & W/Rec/PC   & No             & 2                           \\ 
2014/05/24  & DX1  & DC50/V          & Cam/LP & Yes      & Yes         & UCM+FO      & W/Rec/LX7  & No             & 1                           \\ 
2015/12/13 & D17         & DC50/S/Rb & Gem           & No      & Yes         & D+UCM         & Sony A7S                 & Pas  & 3                           \\ 
2016/01/04 & D18         & DC50/S/Rb & Qua        & Yes      & Yes         & D+UCM         & Sony A7S                 & Pas         & 1                           \\ 
2016/04/23 & D19 & DC50/S/Rb & Lyr             & Yes      & Yes         & D+IAA         & SA7S/Go4    & Pas         & 3                           \\ \cline{1-10}
\end{tabular}
\end{center}
\end{table}

\subsection{Planning}
\label{sec:plan}  

A mission usually starts weeks ahead the launch. Planning includes the design of the flight, the integration of the craft and instrument, and the request of flight permit when needed. 

Main constraint for the design of the flights is the recovery in a easy-access area, far from the seaside. The launch location is always in the center of the Iberian Peninsula, where part of the team is located. Other members of the team are based on locations from NW to SE Spain, consequently the logistics for the assembly are complex. 

The flight is simulated following a protocol including different scenarios (balloon lift $\pm$ 10\%, premature balloon burst, payload weight $\pm$ 10\% and others). The implementation of the evaluation of the scenarios is one of the lessons learnt from some difficult recoveries. 

\subsection{Launch, lift and landing}
\label{sec:lift}  

The launch is the longest stage of a mission. It includes the countdown for the release of the balloon and the testing of all systems, using checklists and protocols inspired by the ones of the European Cooperation for Space Standardisation (ECSS). 

The lift is provided by commercial weather balloons, from 400~g to 2000~g. The nominal mass of the probe is 4~kg, and the balloons are filled by 9~m\textsuperscript{3}, yielding a lift slightly over 7~kg. Therefore the standard mission reaches a burst-altitude of 31~km after 80 minutes.

After the burst the probe enters in free-fall phase. After few seconds the parachute opens and starts breaking the assembly as the atmosphere gets thicker.

One of the current developments to increase the control of the flight is the addition of a "cut\&down" system, that cuts off the balloon rope. Probes Daedalus 18 (D18) and 19 (D19) were equipped with this system. The Wolfram filament was damaged in D18 and it didn´t work. For the D19 it worked well, but the cut rope got rolled with the parachute, not releasing the balloon. We are currently working on a new design to solve this problem. 

\subsection{Stabilisation system}
\label{sec:stabilisation}  

We have used the data from magnetometers and accelerometers to study the rotation of the probe. It depends on the configuration of the balloons and the mass distribution within the craft. We consider it stable when the rate is under 1 rotation per minute (i.e., 6 degrees per second). For the usual video exposure time of 1/50\textsuperscript{th}~s, that is less than 7 arcminutes per frame, similar to the plate scale for the high-sensitivity low-resolution meteor cameras. However rotation rates could go over 5 rpm, especially in the troposphere.

For the Geminids 2015 (Daedalus 17) the probe was equipped with active stabilisation using gyroscopes. Unfortunately the results are inconclusive because the batteries ended before the balloon reached the stratosphere.  

  \begin{figure} [ht]
   \begin{center}
   \begin{tabular}{c} 
   \includegraphics[height=5cm]{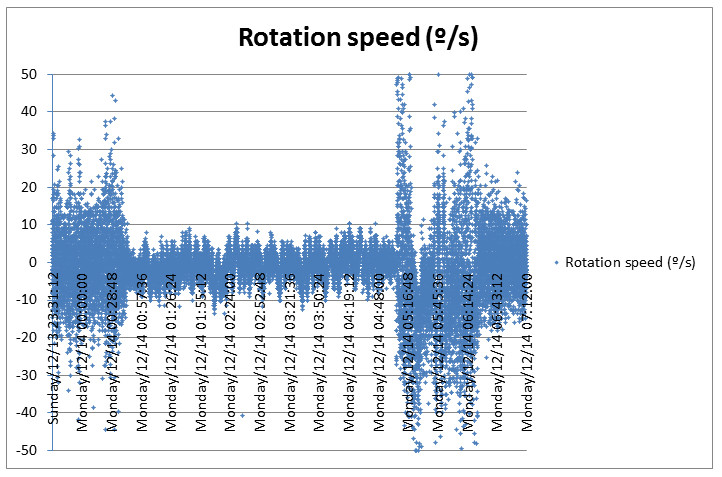}
   \end{tabular}
   \end{center}
   \caption[rotation] 
   { \label{fig:rotation} Rotation rates for the Geminids 2015 mission. The 3 stages of the mission are easily identified, with an unstable tropospheric ascent, a long\footnote{One of the balloons exploded prematurely and the lift was dramatically reduced} pleasant stratospheric flight and a wobbling free fall after the balloon burst.}
   \end{figure} 
   
\begin{figure} [ht]
   \begin{center}
   \begin{tabular}{c} 
   \includegraphics[height=5cm]{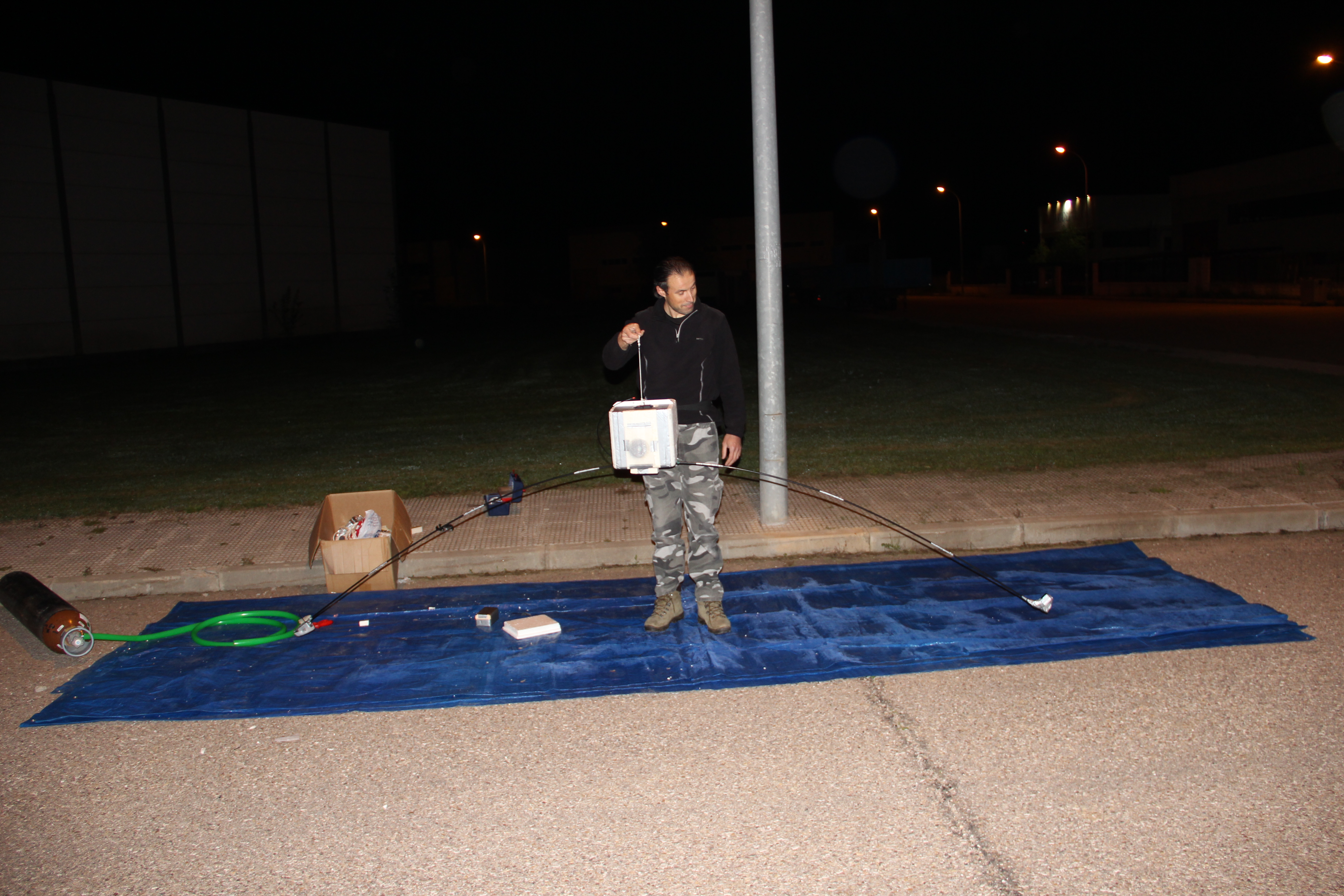}
   \end{tabular}
   \end{center}
   \caption[daedalus19a] 
   { \label{fig:daedalus19a} Probe configuration before launch of Daedalus 19, ready to attach the balloons. The probe has a long rod and the mass is distributed to increase the moment of inertia.Photo credit: Phillip Maier.}
   \end{figure}    

\subsection{Tracking and recovery}
\label{sec:tracking}  

The default mode of the probe is uncontrolled, with unidirectional communication. This option was chosen after several designs in order to keep the simplicity. The use of bidirectional communication was tested in some missions but the ground segment needs more operators and more hardware. Currently it is only considered the option of active control of the probe in the case of the "cut\&down system".

In the design of the system the option of downloading the data was discarded due to the nature of the experimental setup. For meteors and light pollution studies, the data consists on videos from 18MB/s up. That download bandwidth was not affordable with standard RF equipment, and with the cost or mass budgets for the probe and the tracking team.

The probe is currently equipped with different trackers and beacons. We rely on different networks to retrieve the position of the probe. The nominal tracking system is based on Global Navigation Satellite Systems. Currently we are using the American Global Positioning System (GPS) and the Russian GLObal NAvigation Satellite System (GLONASS). They have dual use, and their capabilities are constrained for civilian use in a range of speed, acceleration and altitude. The use of both networks allow us to keep track of the whole mission. The position is sent via satellite communication and served in the manufacturer website.

For recovery we use the GSM network that is usually available in the zone of our launches. The transmitter is a smartphone equipped with a GPS receiver. The phone runs a software\footnote{Strato Beacon Balloon, developed by Aitor Conde.} that broadcasts the position when the probe is under 3000 meters, covering the launch and landing; the two critical phases when the probe could get lost. 

Most of the missions include a backup system: an RF tracker that provides distance and direction of the balloon to the receiving station during the whole missing. This system proved to increase the payload recovery chances, as it happened for the Geminids 2015 failed mission. It was recovered 10-miles off coast, after spending 2 hours without any GNSS signal.

  \begin{figure} [ht]
   \begin{center}
   \begin{tabular}{c} 
   \includegraphics[height=8cm]{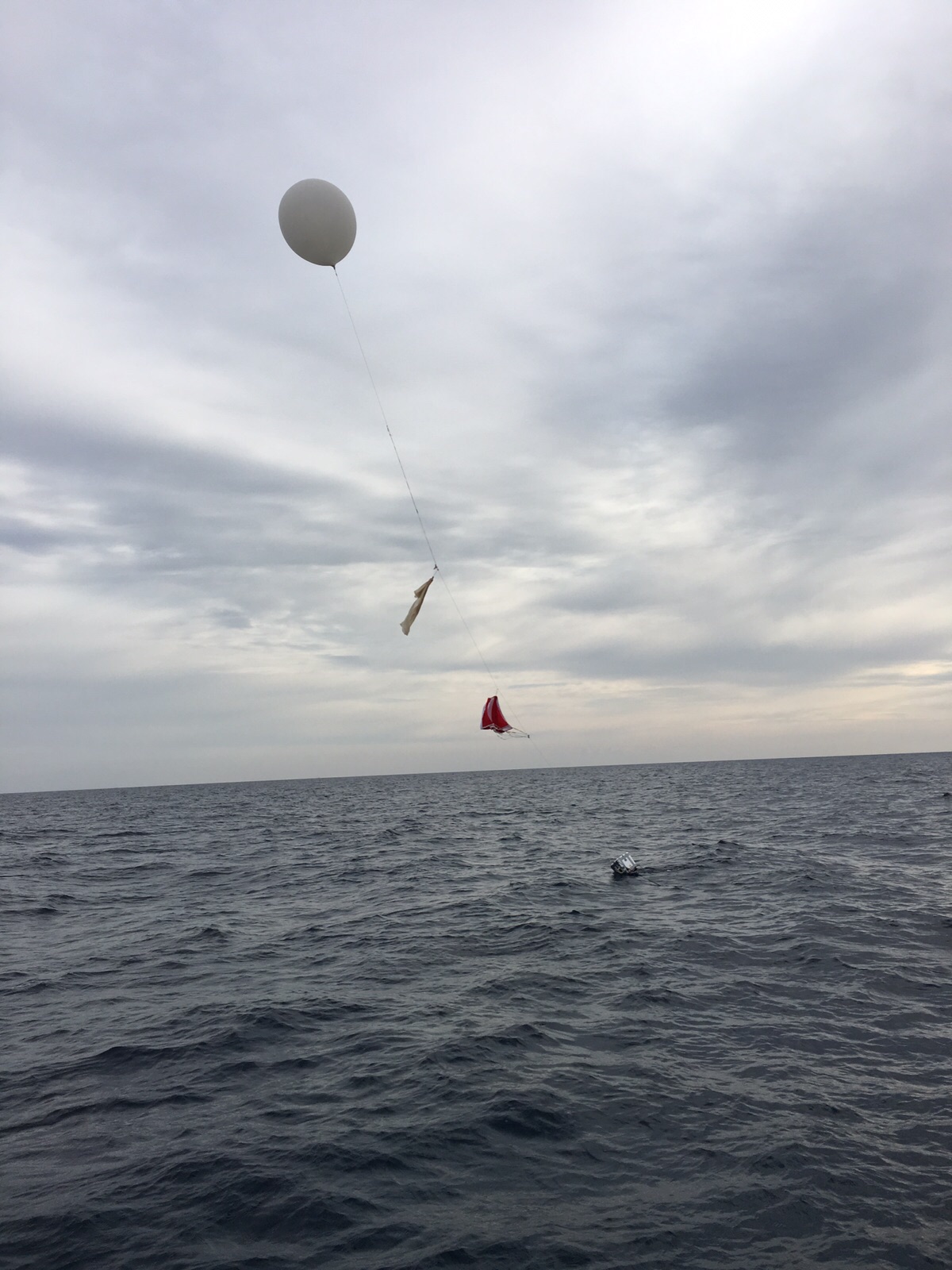}
   \end{tabular}
   \end{center}
   \caption[stars] 
   { \label{fig:stars} Geminids 2015 probe was recovered 10-miles off coast thanks to the GPS and RF beacons.}
   \end{figure} 

The recovery team has always at least two people, the driver and the operator. The operator receives positioning through satellite/internet, GSM and RF receivers. In case the probe lands in an area where the recovery is difficult, the systems may run out of battery before the team arrives. Then the probe is searched during the following day. The probe has always a tag with contact information for the fortuitous discoverers.

\subsection{Recording system}
\label{sec:recording}  

As explained in previous sections all the data generated by the instruments is stored locally and not uploaded to the tracking station. In the first missions we used a videorecorder but the compression algorithm constrained the scientific use of the images. For the mission D4 we used a notebook but the system failed out of the stratosphere as the rare air caused high temperatures in the processor. For the mission D12 we increased the radiative heat dissipation, but the computer was switched off accidentally before reaching the stratopause.

The introduction of new detection systems (cameras Lumix LX7 and Sony A7S) changed the approach as the instruments had storage for the data taken during the mission. 

\subsection{Detection system}
\label{sec:detection}  

The first instruments on-board were B\&W high-sensitivity low-resolution videocameras, like the Watec 902H2 Ultimate that reached stars up to magnitude 2 in very wide fields. 

The last instruments tested in our wide-field platform are Electronic Viewfinder with Interchangeable Lens (EVIL) cameras. The best result has been obtained with the Sony A7S, with a backlit sensor recording full HD colour frames, at 30 fps, with 1/50~s exposure time and ISO~20000. It recorded up to magnitude 6.5 for stars, or 21.5 magnitudes per arcsecond\textsuperscript{2} for sky brightness. 

   \begin{figure} [ht]
   \begin{center}
   \begin{tabular}{c} 
   \includegraphics[height=8cm]{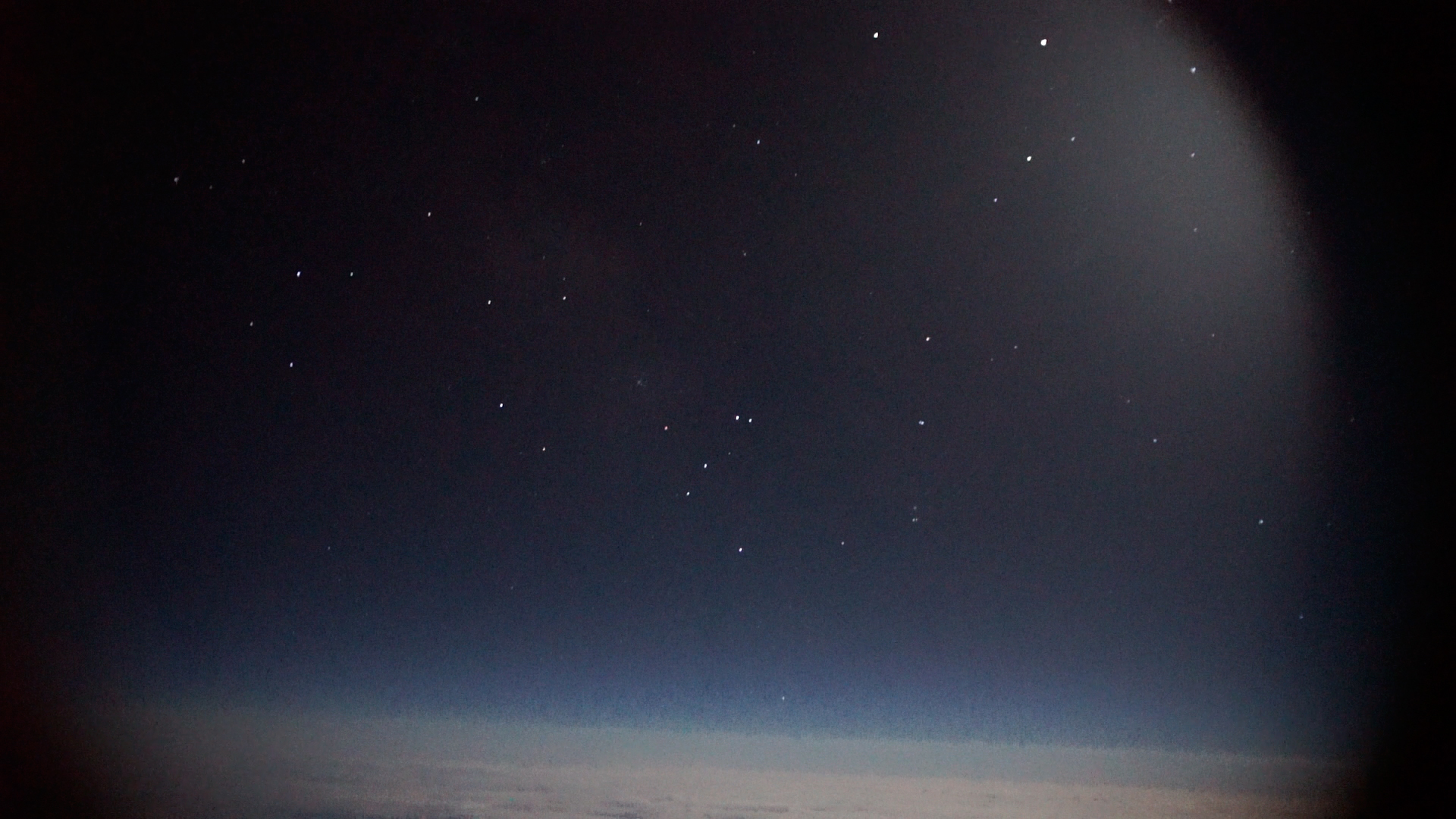}
   \end{tabular}
   \end{center}
   \caption[stars] 
   { \label{fig:stars} Sony A7S imaging Scorpio and Sagittarius constellation, showing stars up to magnitude 6.5. Galactic center is also visible. }
   \end{figure} 

As important as the sensor, the optics are chosen to be fast (low f/ number) and covering the full sensor. For Quadrantids 2016 the lens used was a Sigma 30 mm f/1.4, producing slight vignetting in the corners of the full-format chip. The resulting images had a plate scale of 3.5 arcmin/pixel, and the chip covers a field of view (FoV) of 110 degrees by 60 degrees.

\subsubsection{Control of the detection systems}

For the first missions the detection system was activated during the countdown before the launch, but that reduces the lifetime of the mission.

In the last three missions the camera Sony A7S was controlled using two different systems. For the Geminids 2015, the camera was controlled using WiFi communication messages sent by a Raspberry Pi (a credit card-sized board computer). However the WiFi suffered from interference and the system failed. For Quadrantids and Lyrids 2016 we used a cable trigger commanded by an Arduino micro-controller.

\section{Scientific output}
\label{sec:science}  

The scientific payloads for these stratospheric flights were focused on light pollution and meteors research. The first of these nighttime flights, Daedalus 2, took place in 2010 as an instrumental test to monitor light. The same year a second flight over the city of Ja\'en successfully took the first images for scientific use\cite{sanchez_de_miguel_2011_52562}. In 2014 the project was repeated over the city of Madrid, acquiring images of quality comparable to the ones from airborne instruments with a hundredth of the cost.

The other scientific objective of these night flights is the estimation of meteoroid influx at Earth through the observation of meteor showers. Observing meteors from the stratosphere improves detection efficiency thanks to much lower extinction and less background brightness. Other teams had already demonstrated the benefits of airborne observations for these objects. Thanks to the low cost of balloons and high-sensitivity cameras there is a number of groups using stratospheric balloons to record meteors\cite{moser20132012,koukal}.

\subsection{Fireballs and meteors}

In 2011 the team sent a low-light video camera to observe the Draconids 2011 outburst\cite{ocana2013first}. Unfortunately the recording system failed and the experiment had to be redesigned. For Geminids 2012\cite{sanchez20132012,sanchez_de_miguel_2013_52570}, Camelopardalids 2014 and Geminids 2015 \cite{sanchez_de_miguel_2016_52574} the instruments have worked flawlessly and the data are under analysis. Imaging and video devices with broad-band filters have been employed and results are comparable with airborne and satellite products, especially for high-resolution images\cite{kyba2014high}. 

For Quadrantids 2016 the instrument was the Sony A7S, recording full HD colour frames, at 30 fps, with 1/50 s exposure length and sensitivity of 20000 ISO. This colour information provides basic spectral information from the objects\cite{ocana2012narrow}. Unfortunately the mission ended prematurely due to balloons malfunction. Preliminary results confirm that 12 QUAs were recorded in 7 minutes\cite{sanchez20162016,sanchez_de_miguel_2016_52630}.

In April 2016 the authors launched the last mission to date to record the Lyrids annual meteor shower, and we are currently analysing the video data\cite{sanchez_de_miguel_2016_52775}.

\subsection{Light pollution research}

Nighttime Earth remote sensing using satellites \cite{kyba2014high}, planes, unmanned aerial vehicle (UAV) or balloons is a growing research are because of its applications in light pollution control, energy efficiency measure, socioeconomic studies, environmental impact, etc \cite{sanchez2015variacion}. The use of balloons for these purposes overcomes the legal limits in the use of other alternative systems like UAV/drones. 

The images taken during the flights performed to date have been analysed in the thesis of ASdM\cite{sanchez2015variacion}. The flight over the city of Madrid produced images with a resolution down to 2 meters per pixel, and proved the feasibility of the technique to measure ground light levels and do multispectral imaging for light pollution sources characterisation.  

   \begin{figure} [ht]
   \begin{center}
   \begin{tabular}{c} 
   \includegraphics[height=8cm]{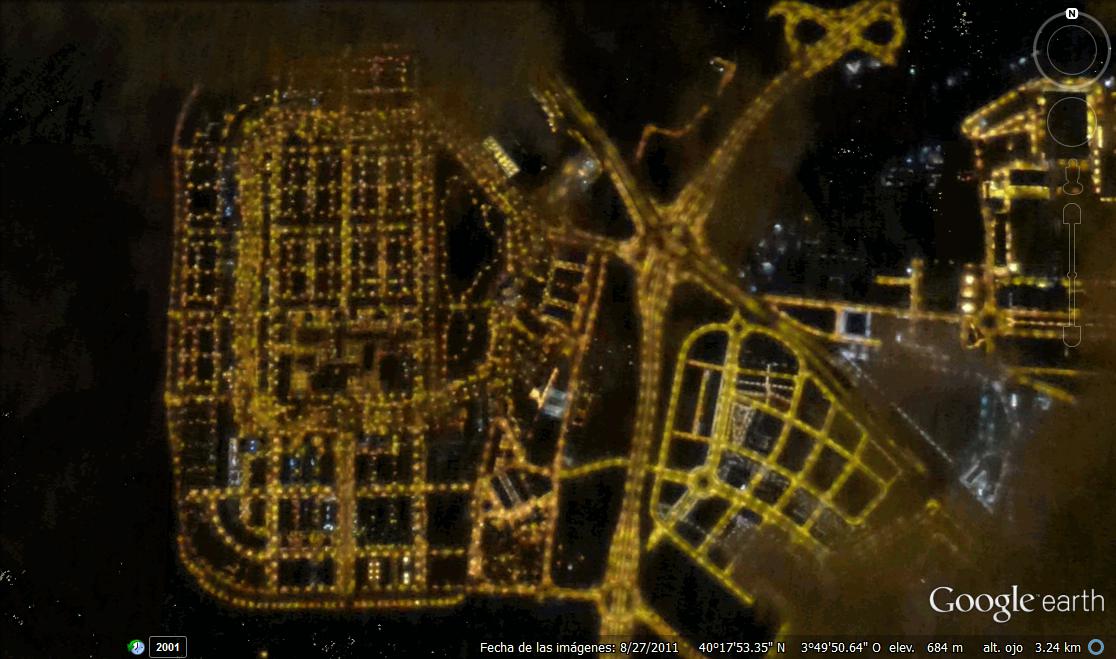}
   \end{tabular}
   \end{center}
   \caption[Lx7] 
   { \label{fig:Lx7} The camera pointing to the nadir during the Camelopardalids 2014 mission took images that allow resolving individual light sources. The images are georeferenced and published as KML files with geographic information system (GIS) to superimpose onto a 3D globe, for instance Google Earth.}
   \end{figure}

\section{Future missions}
\label{sec:future}  

There area some missions scheduled for the rest of 2016 in order to keep this successful Pro-am collaboration. There are two scientific missions foreseen for Perseids and Geminids, and an experimental one to test heavier optical elements (including a spectrograph).

In the medium term the technological efforts are focused on improving the platform stability, increasing the payload mass and increment the scientific output of the missions. In the medium term, this launches will eventually join the ORISON missions, intended for payloads between 15~kg and 50~kg. 

\acknowledgments 

The authors want to thank the support from the GUAIX instrumentation group of the Universidad Complutense de Madrid, and to all the students and researchers that have participated in the different missions: Mario Fern\'andez Palos, Jaime Izquierdo, Carlos E. Tapia Ayuga, Cristian V\'azquez, Sandra Zamora and Jaime Zamorano. Also to Phillip Meier tha came on the last mission.

The research leading to these results has received funding from the European Union’s Horizon 2020 research and innovation programme under Grant Agreement No. 690013.

Special thanks to Daedalus Team current and former members: Pedro Le\'on, Fernando Ortu\~no, Aitor Conde, David Mayo, Miguel \'Angel G\'omez, Rub\'en Raya, Mar\'ia Teresa Ortu\~no and Alejandro S\'anchez de Miguel. 

AstroInnova thanks to all the partners that have sponsored some of the missions.

\bibliography{report} 
\bibliographystyle{spiebib} 

\end{document}